# Carbon Memory Assessment

## Franz Kreupl

Technische Universität München, Department of Hybrid Electronic Systems, Arcisstr. 21, 80333 Munich, Germany, franz.kreupl@tum.de

## Introduction and Background

This is a white paper for the ITRS meeting on emerging research devices (ERD) in Albuquerque, New Mexico, on August 25-26, 2014. The geometrical and performance scaling of silicon CMOS integrated circuit technology over the past 50 years has enabled many affordable new products for business and consumer applications. Recognizing that Flash is approaching its ultimate physical scaling limits within the next 10 years or so, the global electronics research community has begun an intense search for a new paradigm and technology for extending the functional scaling of memory technologies. Several promising nonvolatile memory concepts have emerged, based on different switching and retention mechanisms from those of Flash memory, e.g., STTRAM, RRAM, PCM and more recently, *resistive memories based on carbon, which are the topic of this paper*. This paper will introduce into the diverse field of carbon materials by recollecting some effects in carbon that can be used to produce a multiple time switchable, non-volatile unipolar resistive memory with potential high scalability down to atomic dimensions. Carbon-based memory is a non-volatile resistive memory, therefore, the same architectures, circuits, select transistor or diodes like in RRAM or PCRAM can be considered as implementation. The big advantage of carbon memory might be the high temperature retention of 250 C, which makes it attractive for automotive and harsh conditions.

## Basic Facts about Carbon

In order to understand the switching mechanism in carbon-based resistive memory, background information is given in this chapter that might facilitate more insight into the complex behavior of carbon.

### Carbon Allotropes:

Over the last decades a plethora of new structures formed by elemental carbon have emerged. Besides the classical form of carbon like diamond and graphite, well ordered formations like graphene, nanotubes or fullerenes and less ordered structures like carbon nanofoams, nanoporous carbon, amorphous carbon (a-C) or diamond-like carbon (ta-C or a-C:H), which additionally can include hydrogen or nitrogen, have been discovered and studied. The electronic properties of these materials can range from metallic to semiconducting to insulating, while the mechanical behavior cover everything, from soft to very hard. The mass density of these carbon structures can vary by orders of magnitude, from 4 $mgcm^{-3}$ for nanotube-based aerogels [1] to 0.2 - 1 $gcm^{-3}$ for carbon nanotube bucky paper [3], 2.2 $gcm^{-3}$ for graphite [3] and 3.5 $gcm^{-3}$ for diamond [3]. For all these different carbon formations, graphite is the most stable form and all other forms, especially $sp^3$-based bonds in diamond and ta-C favor the relaxed $sp^2$-bond [3].

### Manipulating the Bond Structure of Carbon:

There is ample evidence that external forces can influence the detailed carbon bond scheme. The conversion from amorphous or $sp^3$-bonds to $sp^2$-bonds may be initiated by stress relaxation [3] and external energy like, e-beams [4], electrical current [5, 6], laser pulses [7, 8], x-rays or temperature [9, 10]. Graphitization without the help of a catalyst usually happens at temperatures above 2500 K, but it has been shown that a relatively small current can heal defects in nanotubes and graphene and even lead to the transformation of amorphous carbon into $sp^2$-type carbon [5, 6]. The reason for this can be seen in the power that is delivered to the object by driving current through it. The power per volume is given by $j^2 \bullet \rho$, with the current density j and the specific resistivity of the material ρ. Even a small current of 100 nA, if focused on a filament with 1 $nm^2$ cross-section, can yield current densities in the order of 10 $MA/cm^2$. As a consequence, the delivered power density is in the range of $MW/cm^3$ to $GW/cm^3$, depending on the resistivity. This provides enough local energy density to initiate the transformation of amorphous carbon to a more $sp^2$-bonded form of carbon. Similar annealing effects have been observed for laser pulse treated carbon. If the energy density of the laser exceeded a certain threshold the formation of an amorphous carbon phase has been observed [8]. In this context, carbon behaves the same way like chalcogenide phase change materials used in read/writeable CD and DVD disks. A short, high-energy laser pulse creates an amorphous state whereas a longer laser pulse with lower energy leads to crystallization. Transformation of $sp^2$-bonded carbon to $sp^3$-phases has been observed by simply changing the amount of stress in the carbon layer [11] or by applying current [12]. Even single-walled carbon nanotubes are locally modified to different carbon structures by high current densities [13].

**Sustainable Current Density in Carbon Structures:**

The maximum sustainable current density in predominantly $sp^2$-bonded carbon structure ranges between 10 $MA/cm^2$ to 1 $GA/cm^2$, which is much higher than what ordinary metals like copper can tolerate [13, 14, 15, 16]. The evaluation of the effective, current conducting area needs to be considered in low mass density structures, which can have a lot of nanoscale voids (like nanotube ribbons or bucky paper), as the current might be focused in channels whose cross-section is different from the apparent outer dimension of the conducting element. The current density can induce very high local temperatures by the associated power density $j^2 \bullet \rho$. This high temperature stability is utilized when using carbon as crucibles in e-beam evaporation systems. On the other hand, care needs to be taken if a carbon structure is operated at its maximum current density at a longer time, as it will destroy and evaporate almost any other material, which is in contact with it [13, 17]. If such a localized high current density is in contact with a metal, it will induce melting and diffusion of the metal [18]. The effect can be used to "photograph" the diameter of filaments in carbon [14]. If current pulses should manipulate carbon, it is important to impose a time limit on the duration (pulse length) of the current flow, as the generated heat in short pulses is focused in a very localized point. If a filament is formed in a-C:H materials the heat in the filament will lead to outgassing of the hydrogen and as a consequence a shrinkage of the material is occurring [19].

**Break-junctions and Plumbing in Carbon Structures:**

When the maximum sustainable current density is exceeded, deconstruction by carbon evaporation is observed in graphene [20, 21] or graphitic stripes [22, 23] or nanotube structures [13, 24] which are in an open environment, i.e. the structures are not embedded by a protecting insulating layer but are held in a vacuum environment like a probe chamber or SEM or TEM. Consequently a very narrow gap is created in these carbon structures, which creates a high resistance path. The creation of the gap by dc-currents lead to relative wide gaps in the order of <10 nm. Short electric current pulse destruction yields smaller gaps. Interestingly these nano-gaps show switching to conductive behavior if a high enough electric field is applied [20, 21, 22, 23, 24]. Due to the experimental situation of the sample in a vacuum environment, the switching is believed to be due to vacuum background contamination by hydrocarbon gas. Even at a vacuum pressure of $10^{-7}$ mbar, a molecular density of ~ $3 \cdot 10^9$ molecules/cm$^3$ is present in the vacuum, albeit not all of the molecules are hydrocarbons. This background gas is also responsible for a-C deposition in SEM samples during scanning with the electron beam [25] and trapping of carbon contamination by an applied electric field is a well know problem in nano-relays [26]. The electric field polarizes the gas molecules in the nano-gap and the current pyrolizes the hydrocarbon molecules and forms a carbon bridge over the nano-gap [24]. This interpretation is substantiated by the observed temperature dependence of the effect. If the sample stage is cooled down, it effectively acts like a cryo-pump and freezes out the gas molecules in the vicinity of the sample [25], thus lowering the probability that gas molecule can be trapped by the local electric field. Due to the open and hard to be controlled environment, this type of switching is less relevant for memory applications, but it can be instructive to learn, that switching in a nano-gap is possible.

However it is more than relevant that carbon structures, which are separated by a small gap, can be plumbed together by the application of an electric field and subsequent current flow. It is has been shown that single-walled [27] and multi-walled carbon nanotubes [28, 29] as well as graphene [30, 31], can be joined and welded by applying an appropriate current flow in the carbon nanostructures. Joule heating originating from the field emission current in the nano-gap will cause atom diffusion and rearrangement of carbon. Of course, this would "close" a nano-gap and lead to a low resistance value. The amount of current for welding and plumbing is quite small and ranges between 0.5 µA to 10 µA [27, 28, 29].

The break-junction mechanism is also likely to occur in low mass density materials like nanoporous carbon, carbon foam or nanotube ribbons, which are encapsulated by a passivation layer. Once the critical current density is reached in a carbon constriction, the carbon simply will evaporate into the local mini-vacuum-environment created by the pores and condense on the carbon structure again [32, 33]. As the cross-sections of bridging carbon structures is very small, currents in the order of 10 µA per contact junction will induce high enough temperature for evaporation. For high mass density material the option of break-junction forming is not available. The thermal shock wave in the high mass density material will lead to a transformation into amorphous or more sp$^3$-bonded carbon [11, 34, 35].

## What Is Not Considered as Carbon-based Memory

There are several studies out, which are diffusing metal ions into insulating phases of carbon to form a resistive memory effect based on metal filament creation and annihilation [36, 37, 38, 49]. This form of resistive memory is not considered here, due to high current operation and hence limited scalability. To the best of my knowledge and experience, metal diffusion occurs in almost all situations (like in ref. [37, 49]) where the capacitance discharge current from the first forming event is done by dc-voltage sweeps on samples with no on-chip current limiter, like on-chip resistors or transistors. In this context it is important to note that a compliance current limit in source-meter or semiconductor analyzer can neither limit the capacitance discharge current nor a dc-current on a time scale shorter than ~30 µs (depending on the brand of the source-meter). As a consequence a very high capacitance discharge current is flowing, followed by a < 30 µs imprint current, given by the series resistance and the applied forming voltage. The filamentary nature of the current leads to very high current densities that dissolve metal into the carbon material. The dissolved metal forms an open or closed filament that represents resistive switching in this case and the whole device works more or less like a CBRAM memory.

In addition, there are reports that show electronic memory effects in insulating forms of carbon films. The injected charge carriers modulate the tunnel barriers and therefore change the current flow, but they show volatile behavior and or not considered here [50, 51].

## Carbon-based Resistive Memory Phase Diagram

Based on the basic facts about carbon discussed above, two different carbon memories can essentially be proposed, depending on the carbon material mass density. The situation is depicted in Fig.1. If the mass density of the material is low, a break junction will form upon the application of a current pulse. This will interrupt current conduction and a high resistance path is created. Upon the application of a voltage pulse field emission current in the nano-gap will cause atom diffusion and welding and closing of the nano-gap, which in turn will form a low resistance path for electron conduction. If the carbon mass density is high, resistive switching will occur in a transformation between $sp^2$-bonded and amorphous carbon. Apparently there must be a transition region in the mass density where both form of switching can occur. The precise value of the mass density is presently not known, but an estimate is indicted in Fig.1 as an axis break. The color-coding employed in Fig.1, illustrates the required current density to induce a resistance change in qualitative way. To put it simply: the more conducting bonds per cross-section area there are, the more current is required. Therefore, the current density increases heavily if high mass density carbon needs to be transformed. The current density of 400 MA/cm$^2$ is the upper limit, measured in graphitic structures whereas individual carbon nanotubes can go up to GA/cm$^2$. The amorphous region can be switched back to more conductive $sp^2$-bonds by applying an electric field that induces electrical breakdown of the amorphous region.

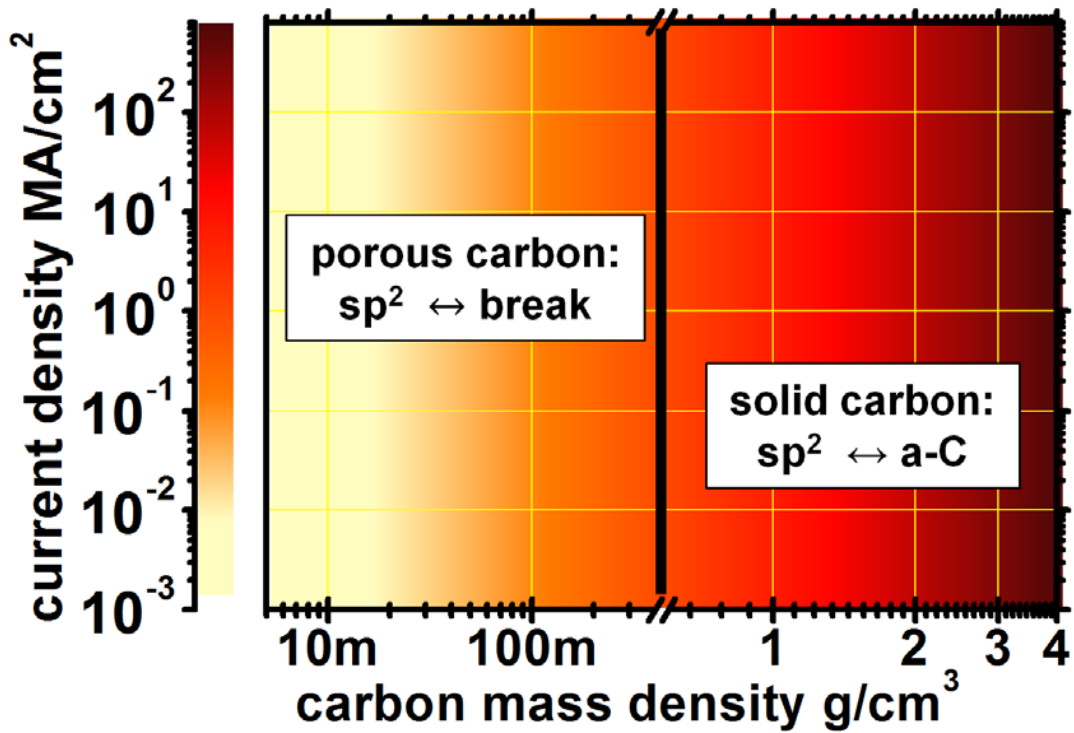

*Figure 1: Proposed phase diagram for resistive memory switching in carbon. If the mass density is low, nano-gap formation will occur upon the application of a high enough current pulse. In high mass density materials a transformation between $sp^2$-bonded and amorphous carbon is proposed. The required current density is color-coded into the image in a qualitative way. In essence, the higher the areal conducting bond density between carbon atoms, the more current is needed to induce a change in the bond structure.*

Porous carbon material can be formed by various methods. Carbides, like SiC or TiC, can be deposited and a subsequent chlorine etch will give a porous carbon structure, which can achieve up to 80% voids in the volume. Alternatively a-C:H with high hydrogen content can be deposited with plasma enhanced CVD and encapsulated. An annealing step will outgase hydrogen and leave a low mass density porous structure. Volume shrinks of more than 60% can be achieved in such a-C:H films, depending on the used deposition process. Carbon nanotubes, particles or fullerenes may be spin coated onto wafers in varying thickness and further processed. For high density carbon layers, a conductive a-C:H with very low hydrogen content can be deposited by plasma-enhanced CVD. Most integration scenarios exclude the direct deposition of graphenic carbon layers as they require temperatures above 860 C, which is too high for backend integration. For very high mass densities, carbon in form of an insulating ta-C can be put on a wafer by using C-ions generated from a dual-bend filtered cathodic arc deposition system. Alternatively, binary insulating materials like sputtered SiC-layer might be used. Bridging of $sp^2$-bonds might also occur during electrical breakdown in amorphous SiC-films.

## First Current Pulse Challenge with Conductive Carbon

If a memory cell is made with carbon material that is in a conductive state right after the fabrication, the first current pulse needs to
  (i) break all the bonds in low mass density carbon or
  (ii) amorphize the whole cross-section of the conductive carbon in high density materials.

In both cases, the current flow will initially reduce series resistance effects before a break or amorphization will occur. As it has been pointed out in the introduction, the sustained current density in carbon is very high. In fact it is more than an order of magnitude higher than what a select device like a diode (10 $MA/cm^2$) can provide. Consequently, if it is assumed that the cross-section of the memory cell and the current-carrying cross-section of the select device are equal in size, it is only possible to break a low density carbon material – a conductive high density carbon material would require more current than the select device can deliver. As a workaround for this challenge, sub-lithographic feature sizes for the carbon memory cell have been suggested for high mass density carbon [14].

The situation is illustrated in the following Fig. 2. A nanotube felt deposited by spin-coating represents a low mass density carbon in Fig. 2(a). As the nanotubes makes random contacts in this network, a percolation path for current conduction exists between the top electrode and the bottom electrode. In order to create a high resistance path between the electrodes the contact points needs to be destroyed by a high enough current density flowing through them. If the current density is not high enough the current path will have an improved conductivity after the current pulse. Once the current density is high enough, the contacts will be broken and nano-gaps have formed. This is visualized in Fig. 2(a) as a white line. Switching to a state with low resistance (on-state) is then realized at one or more locations, where $sp^2$-bonds bridge the nano-gap again. Further switching requires much less current, as only these individual bonds must be broken. If this first current pulse requires more current than the select device (diode, transistor) can deliver, operation of this type of memory is not possible due to the first pulse requirement.
The variability of cell resistance and random contact junctions will be low for large area memory cells as there is a lot of averaging. Consequently a current density value below 10 $MA/cm^2$, which is manageable by a select device, might be feasible. Severe problems arises once the cell area scales down to the size of the constituents, i.e. size of the nanotubes or pore diameter. At small cell areas the variability of cell resistance and contact junctions in the random carbon felt will be very high, so that the required current density might exceed the values of select device. A solution to this fundamental scaling problem can only be achieved by having well aligned porous materials at the nanoscale.

Even worse is the situation, depicted in Fig. 2(b), where a high mass density conductive carbon is used as the starting material. The first current pulse needs to amorphize the whole cross-section of the memory element in order to go to a high resistance state. The typical current density to achieve this is up to 400 MA/cm$^2$ [14] This would be equivalent to a current of 40 mA for the 100 nm node and still around 400 µA for the 10 nm node. Clearly, this is too much current for any select device and the associated voltage drop on the interconnect wires would be substantial. Only the creation of a sub-lithographic pore-cell would enable a conductive high mass density cell, that is switchable. But the pore diameter needs to be in the range of 1-2 nm to give reasonable low current values [14]. The carbon pore cell may be considered as the "heater" in a PCM mushroom cell, just that the "heater" itself is switching in a carbon cell.

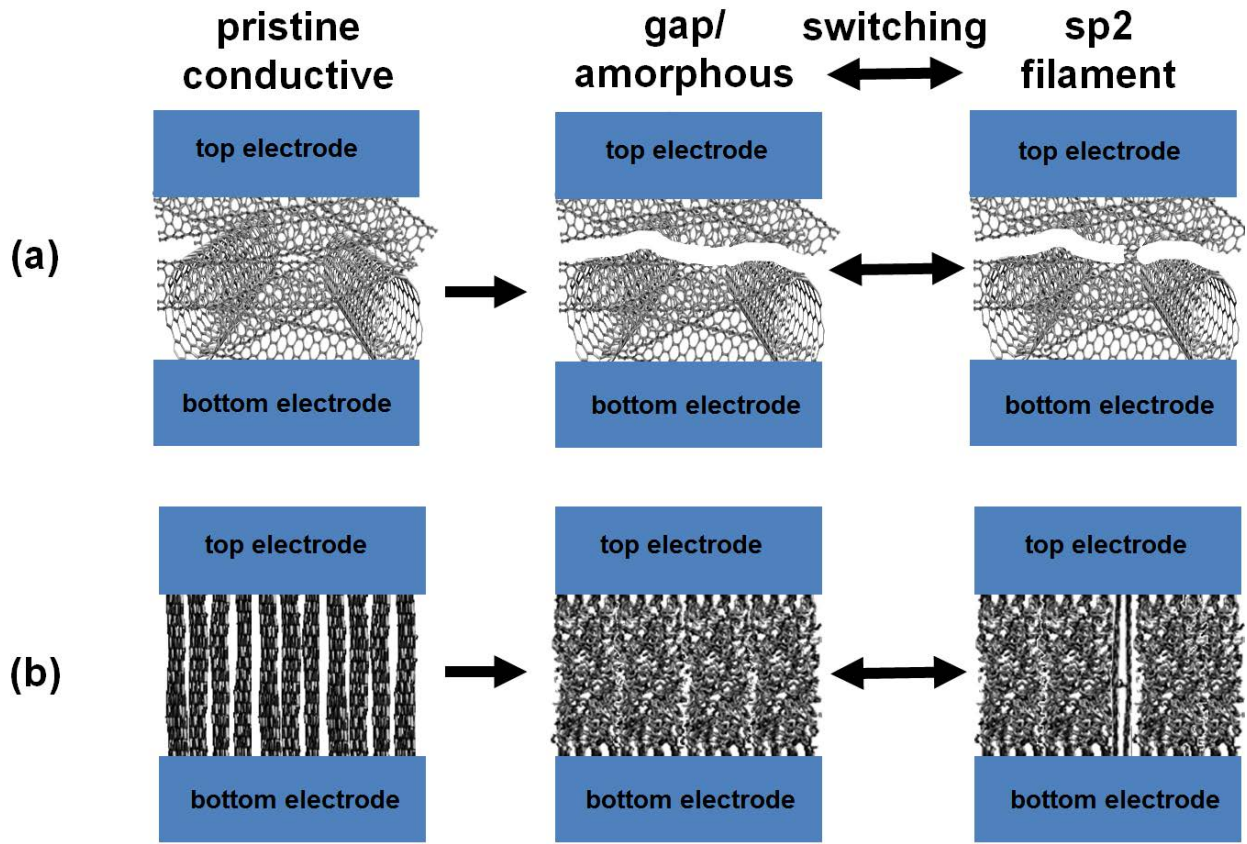

*Figure 2: (a) shows a pristine conductive carbon memory cell based on low mass density carbon nanotube ribbons. After the first current pulse, the percolation-based current path in the random network is destroyed and nano-gaps have formed. For on-switching events one or more sp$^2$-bond bridges will form in the nano-gap. The next reset current pulse needs only to destroy these individual sp$^2$-bridges. (b) illustrates the situation of a high mass density conductive carbon. The first current pulse needs to amorphize the whole cross-section. Subsequent switching events can then be based on individual sp$^2$-bonds and require much less current.*

# Forming Challenge in Insulating Carbon

Seeing the problems of starting with a conductive carbon memory, one can take advantage of the fact that carbon can also be deposited in an insulating form like ta-C, a-C, or a-C:H. In the case of insulating carbon, a forming event like in most oxide-based RRAM memories needs to be performed. The material needs to be biased at the break-down electric field, which depends on the exact nature of the carbon film. The breakdown occurs very fast and as a consequence the capacitive discharge current through the carbon layer, given by $I_{Cap} = C \cdot dV/dt$, is very high. If there is no current limiting on-chip resistor or transistor available which can screen some of the external capacitance, the current can shoot up to 10-20 mA, as a simple tip-probe on a 50 µm diameter contact pad has more than 1 pF of capacitance. Due to the high current density, metals from electrodes will diffuse into the carbon, if the forming is done in a dc-sweep instead of an energy limiting short pulse [36, 37, 49].

In almost all cases of memory cells that need forming, $V_{Form}/I_{Cap}$ roughly gives the on-resistance state of the memory device with thin non-conducting layers, where the forming voltage $V_{Form}$ is given by the voltage value where breakdown occurs and the capacitance discharge current peak is $I_{Cap}$. One example from Dellmann et al. [39] is depicted in Fig. 3. It shows the current response upon the application of a forming voltage pulse with the typical sharp capacitance discharge current peak. In materials with thicker layers, a built-in resistance path of the layer can form an additional resistance that can limit the capacitance discharge current. For the reset current, a current peak that is at least as high as $I_{Cap}$ needs to be employed. Therefor, it is mandatory to limit the capacitance discharge current. A simple approach is to make carbon materials with lower breakdown voltage. Usually this is achieved by a thinner carbon film thickness. Forming voltages smaller than 1 Volt have been demonstrated in dc-forming in diamond-like-carbon films [40, 41], but it is not well known how this translates to pulse forming.

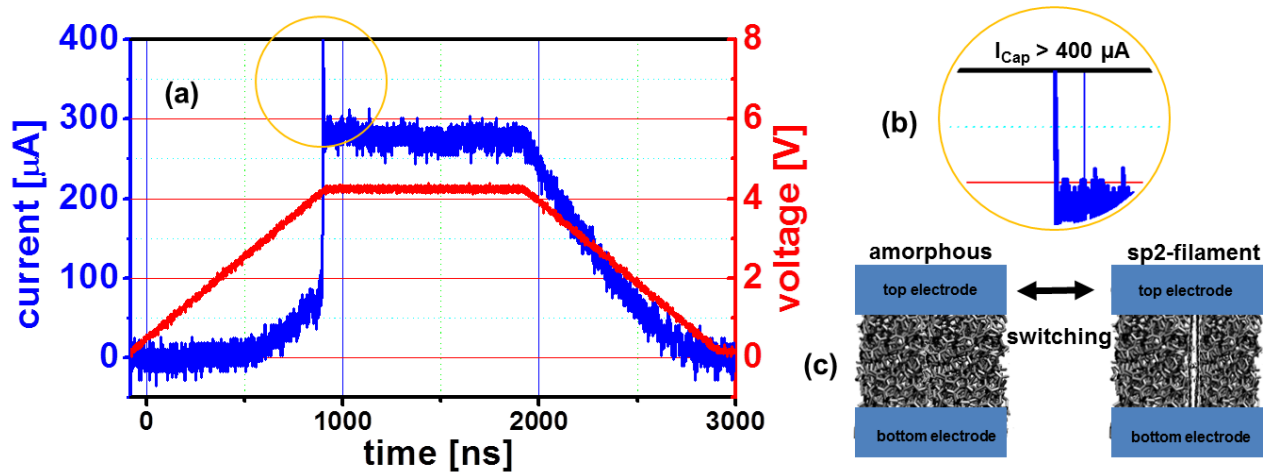

*Figure 3: (a) shows forming voltage pulse and current response pulse of a-C:H memory cell fabricated by Dellmann et al.[39] (data extracted from[39]). (b) is a magnified region of (a) and shows the capacitance discharge current peak, which is at least 400 µA high. The on-resistance of the memory cell of Dellmann et al. is ~ $V_{Form}/I_{Cap}$ ~ 4/400µA = 10 kΩ. (c) shows the schematic of inducing $sp^2$-filaments in an amorphous carbon material.*

## Scaling of Carbon Memory

The dimensional scaling of the carbon memory goes down to single atomic bonds - in both varieties, the low mass density as well as in the high mass density cells. In gap-induced switching cells, single carbon chains will form the bridging and in $sp^2$-to-$sp^3$ conversion cells, one $sp^2$-bond would change the resistance. The structural aspect of scaling is one thing, but the other important information would be, how much current and voltages are needed for operation of this memory cells. Fortunately, there are some TEM-studies out, which have investigated both, the structural as well as the conduction behavior of single chains of carbon [27, 28, 29, 31, 42, 43]. From these studies, the typical voltages and currents can be deduced. Single-walled carbon nanotubes broke at 12 µA and joined again at the application of 1.6 V and 6 µA annealing current (corresponding to a current density of $7 \cdot 10^8$ Acm$^{-2}$). For atomic chains of carbon, currents between 1 nA and 10 µA have been reported with applied voltages ranging between 0.4 and 1 V. Clearly, further investigations are needed in this context to determine the scalability and associated reliability aspects of it.

In the context of scaling, it needs to be addressed that the biggest threat for scaling down (Carbon) ReRAM technologies might lie in the challenge of managing contact resistance between different materials that need to be connected. The relation between absolute current value and current density, which is responsible for material damage, is plotted in Fig.4(a). At the 10 nm node, the delivery of 10 µA current would result in a current density of 10 MA/cm$^2$. The sharp increase of absolute contact resistance with technology feature size is plotted in Fig.4(b). At 10 nm node size, one contact could yield a resistance of 100 kΩ to 1MΩ. The associated voltage drop at the contact, as indicated in the inset of Fig.4(a), is huge, if 10 MA/cm$^2$ needs to be driven through the contact. Typical reported values for metal contacts to graphene or graphite are in the order of $10^{-6}$ to $10^{-5}$ Ωcm$^2$, which is alarming. There is a clear need to study bit-cost-scalable 3D technologies which are not running into the problems of contact resistance.

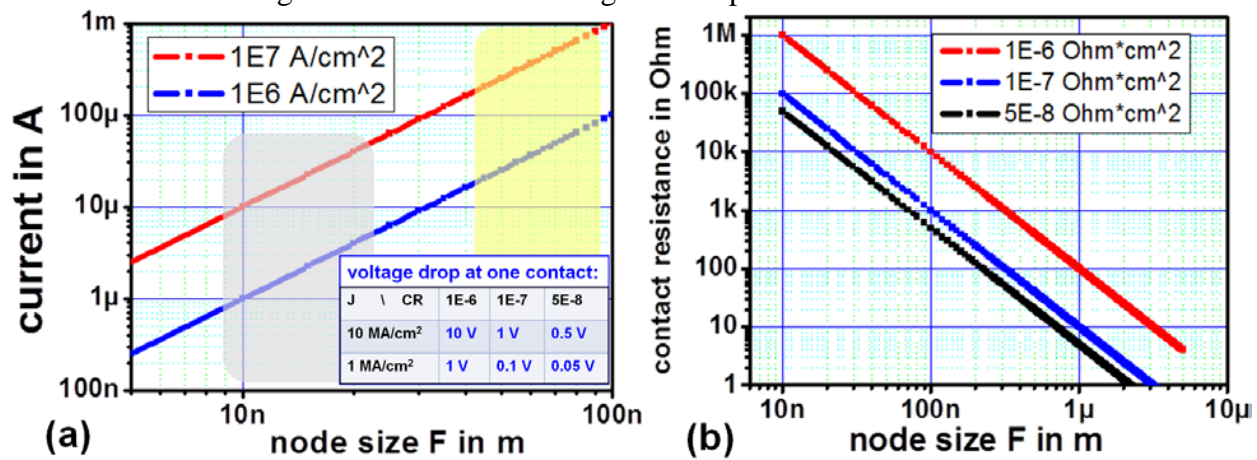

*Figure 4: (a) shows the relation between absolute current value and the current density for a given technology node. (b) demonstrates the increase of the absolute contact resistance value for typical specific contact resistivities as the technology node is reduced. The table inset in (a) illustrates the associated voltage drop per contact for different contact resistivities.*

## Architectural Challenges for Resistive (Carbon) Memories

While the principle operation of resistive (carbon) memories with various materials have been demonstrated in single cells or small arrays with very relaxed feature sizes, the real challenges show up once the memory should be operated at competitive feature sizes and densities. In relaxed scaled memory arrays, the bit line interconnects can be regarded as low-resistance Ohmic conductors, and transistors can be seen as ideal on/off switches. In highly scaled memory arrays the bit-line interconnect wire in connection with contact resistances are huge resistors and the select device is also a reasonable resistor, whose voltage drop depends on the amount of current that is flowing through them. In addition, the serial resistance arising from the interconnect wires, that an individual memory cell has, depends on where it is connected to bit-line. At the very beginning of the bit-line, the resistance might be low, and at the very end, the bit-line resistance is at its highest value. This makes the manipulation of the resistance state of the memory cell very complicated. And the fact that the drop of a pico-probe on the bit-line adds additional capacitance and therefore energy to the system complicates debugging of the memory arrays behavior. In that sense, the effort to gain insight by measuring the behavior (by a pico-probe) changes the relevant physical quantity of operation (the capacitance of the cell environment).

In unipolar memories, the set-voltage is usually higher than the reset voltage (depends on serial resistance). If the memory cell is attached to two serial resistors (one is the bit line, the second is the select device), the voltage across the memory cell rises sharply once the cell has switched to the high resistance state. This is due to the fact that there is now very little voltage drop over the bit-line and the select device, now that the current flow is severely reduced, and all the voltage drops over the memory cell in the high resistance state. The higher voltage over the memory cell makes the cell switch to the on-state again, leading to an unsuccessful reset pulse operation or even several on-off switching events within an erase voltage pulse with unpredictable outcome.

The access points for the circuit designers, to manipulate the state of the memory cell are at the outside of a crossbar memory array. The RC-delay arising from the interconnect wires makes the application of short pulses in the order of 10 nanoseconds a huge challenge. This has severe consequences as discussed above. The quenching of the memory cell requires short pulses, but it is challenging to discharge the bit-line or gates fast after a reset pulse. In PCM circuits additional quench transistors have been proposed, but the exact timing of them is difficult and the RC-delay of them in highly scaled arrays huge. In densely packed crossbar memories with diode-like selector devices, the application of short pulses seems not to be feasible.

A solution to these problems might be available by *operating the memory array in a capacitance discharge mode*. In this mode, the interconnect wires leading to the memory cell are charged up to a predefined voltage level, while the connection after the select device are floating and are not yet connected to ground. The charge that is stored in the interconnect capacitance is used to set and reset the memory cell. After pre-charging, the interconnect wires are disconnected from the voltage supply and the energy of $\frac{1}{2} CV^2$ is stored in the interconnect wires. This is the moment, where the select device is connected to ground and a short current pulse will discharge the charge stored in the capacitance through the memory cell and set or reset the memory cell. This type of operation would allow the application of short current pulses even if the RC-delay from the interconnect wires would impose limitations.

## Current State-Of-The Art For Carbon Memory Technology

The most advanced, relevant data with high statistics, which are available in public domain, are based on publications and the technology of the company Nantero. All other data in public domain are only single cell data with limited statistics. In the spirit of this paper, the operation mode of Nantero's technology is based on low mass-density break-junction formation that is happening in nanotube ribbons and the reported current values for writing and erasing fits well into the proposed mechanism. There are a number of papers out that describe important achievements in this memory technology [44, 45, 46, 47, 48]. The *first pulse challenge*, however, is not tackled, because most of the data are generated on a 4 Mbit test chip in 0.25 µm technology, where the transistor is strong enough to deliver enough current for the first break pulse. The recent publication from Ning et al. [48] used only external transistors, which adds a lot of capacitance and therefore a lot of energy to the operation of the memory cell. However the work of Ning et al. could demonstrate an endurance of $10^{11}$ cycles on individual cells in 140 nm diameter cells [48]. To judge the applicability of this approach for high density integration, the complete current pulse history starting with the first pulse is required. These data are absent or not published up to now.

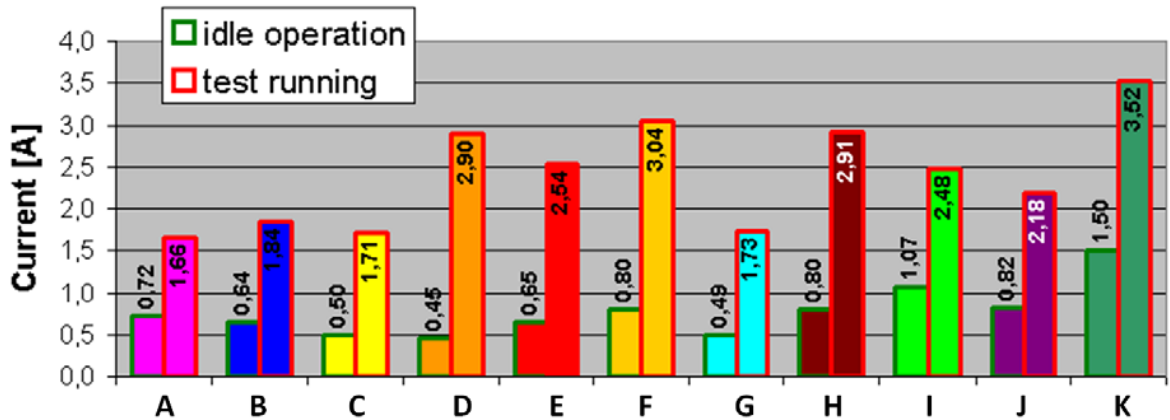

*Figure 5: Measured current uptake from 11 different DRAM chip designs from different vendors (labeled A to K) while being in the idle mode or running the same test program in the same PC. All of them are fully compliant to the same JEDEC specifications, but it is obvious that the individual chip design makes an important impact to the overall energy consumption.*

A first initial assessment of the carbon memory technology is given in the *Appendix 1*. A lot of the questions in Appendix 1 address a certain chip-design and layout. However, all competitive chip designs are company confidential and can of course not be revealed here. Power and energy are always related to a certain bandwidth requirement and cannot be answered without it. As a general rule for energy consumption in memory assessment, it might be worth noting that single cell energies does not reveal too much relevant data, with the exception like CHE programming, where large currents are drained. To illustrate how difficult the judgment of energy consumption even in well known technologies like DRAM, is, Fig. 5 is presented, where the current consumption of different DRAM chip designs from different manufactures are measured in the idle mode and when they are running the same test program. Even in the idle mode, the current uptake can be three times higher in the worst chip design than in the best design. This elucidates that the specific chip design and layouts needs to be taken into account to judge some specification of an unknown memory technology like carbon memory. If data from RRAM and PCM would be available, they would relate well to the carbon memory technology and could be used as a reference.

# Acknowledgement

Wabe W. Koelmans and Abu Sebastian from IBM Research in Zurich provided valuable input and feedback to this discussion.

# Literature and References

# Appendix 1: Metrics for memory device assessment

| | Description | Unobtainable-RAM (e.g., "Ideally…") | CARBON MEMORY | Comments (including any associated tradeoffs) |
|---|---|---|---|---|
| **Scalability, size, cost** | Scalability | Each layer @ 4F$^2$ down to beyond 12nm node | @ 4F$^2$ down to single atomic bonds | both memory cell AND wire pitch can be scaled. But care needs to be taken about the select device. Major threat comes from contact resistances |
| | Multi-level cells (MLC) | Up to 3bits/cell | 2 bits/cell are demonstrated [48], Recommended is 1bits/cell | MLC in RRAM is possible for relaxed feature sizes. IF RC from interconnects play an important role only, Ron > 200 kOhm might be accessible |
| | Multi-layer stacking | At least 32 layers | 8-12 Multilayer might be possible | BEOL compatibility depends on the select device. Complicated stacks needs to be etched at 8-12 layer. 3D monolithic integration might be feasible |
| | Fabrication costs | Total cost very similar to current NAND or lower | Similar to PCM or RRAM | Number of critical mask steps= 1 for 1layer CMOS - No new/difficult unit processes -New/difficult materials only with nanotubes, a-C is known to be compatible with CMOS processing- device forming is necessary (first pulse challenge)! |
| | Array efficiency | >100% (circuitry tucked underneath, w/ extra Si real-estate left over) | Similar to PCM or RRAM | • BL/WL lengths<br>• Extent of peripheral circuitry<br>• Peripheral circuitry play a critical role (such as compliance, or current limiting needed for low power<br>• Interplay with 3D stacking |
| | | | | |
| **State-of-the-art** | Array size | N/A (Unobtaina-RAM has not been demonstrated) | 4 Mbit [45] | obtained individually… |
| | Yield | N/A | Not known | |
| | Technology node | N/A | 20 nm cells demonstrated (in public domain) [45] | **But not for both the implemented CMOS device AND for the wiring pitch (CMOS 0.25 μm) First current pulse issue not investigated** |
| | | | | |
| **Latency (Both cell-level and system-level, if known)** | Read latency | < 10ns (for memory applications)<br><br>(~1us for storage applications) | ~50 ns for 1 [45]<br>~30 ns for 0 [45] | • read contrast ~1000<br>• Size of read window<br>• Read disturb issues<br>• Errors from crosstalk<br>**= all depend on chip design** |
| | Write latency | < 20ns (for memory applications)<br>~1us for storage applications | ~20 ns [48] | • **Requires verify-after-write/erase**<br>• Write disturb of other devices = research<br>• Damage threshold to avoid? = research<br>• write-in-place supported<br>**= all depend on chip design** |
| | | | | |
| **Power / Energy** | Read power / Energy | < 1/10 scaled DRAM for memory applications<br><br>(for storage applications,same as scaled NAND or | 1V/10 nA<br>1V/10μA | Power → parallelism → bandwidth<br>Roadmap with scaling<br>Please specify power usage…<br>• …by selected devices<br>• …elsewhere in the array (leakage, line resistances), and<br>• …in peripheral circuitry<br>**= depends on proprietary chip design…** |
| | | *better)* | | |
| | Write power / Energy | for memory applications, < 1/5 scaled DRAM *(for storage applications, <5x read power)* | 15pJ/bit [48] | Power → parallelism → bandwidth Roadmap with scaling? power usage… <br>• …by selected devices <br>• …elsewhere in the array (leakage, line resistances), and <br>• …in peripheral circuitry <br>   = depends on proprietary chip design… |
| | | | | |
| Reliability | Endurance | >>1e12 (memory applications) <br><br>(>1e9, storage applications) | >1e11 [48] | • Can device failures be predicted <br>= research subject, scales with node… <br>• Devices fail to what state? <br>= research subject, scales with node… <br>• Do failed devices affect neighbors? <br>= research subject, depends on design <br>• Impact on other characteristics? (e.g., do cycled devices behave differently?) <br>=research subject, scales with node… <br>• Are failures random or clustered? <br>= research subject, scales with node… |
| | Retention | >1 month @ 85°C (memory) <br>>10 years @ 150°C (storage) | >10 years @ 120°C <br>300 min @ 300 C [41] <br>168 h @ 250 C [44] | tradeoffs, between retention & write-speed, or retention & cycling? <br>   = research subject, scales with node… |
| | Variability | Extremely low (e.g., 6$^{th}$-sigma device also meets all specs) | If we would have 6 sigma, we would have a product…. | Intra-device & inter-device – variability & repeatability? <br>   = research subject, scales with node… <br>Porous carbon will have a problem at small dimension |